\documentclass[fleqn,usenatbib]{mnras}

\usepackage{newtxtext,newtxmath}
\usepackage{color}

\usepackage[T1]{fontenc}
\usepackage{siunitx}

\DeclareRobustCommand{\VAN}[3]{#2}
\let\VANthebibliography\thebibliography
\def\thebibliography{\DeclareRobustCommand{\VAN}[3]{##3}\VANthebibliography}

\usepackage{graphicx}	
\usepackage{amsmath}	

\newcommand{\XY}[2]{\left[\textrm{#1/#2}\right]}
\newcommand{\FeH}{\XY{Fe}{H}}

\newcommand{\kms}{km\,s$^{-1}$}
\newcommand{\Teff}{T_\textrm{eff}}
\newcommand{\logg}{\log g}

\newcommand{\Msol}{\text M_\odot}

\title[Search for EMP stars in the LMC]{The SkyMapper search for extremely metal-poor stars in the Large Magellanic Cloud}

\author[W. S. Oh et al.]{
W. S. Oh,$^{1}$$^{,2}$\thanks{E-mail: weishen.oh@anu.edu.au}
T. Nordlander,$^{1}$$^{,2}$
G. S. Da Costa,$^{1}$$^{,2}$
M. S. Bessell$^{1}$$^{,2}$
and A. D. Mackey$^{1}$$^{,2}$
\\
$^{1}$Research School of Astronomy and Astrophysics, Australian National University, Canberra, ACT 2611\\
$^{2}$ARC Centre of Excellence for All Sky Astrophysics in 3 Dimensions (ASTRO 3D), Australia
}

\date{Accepted XXX. Received YYY; in original form ZZZ}

\pubyear{2023}

\begin{document}
\label{firstpage}
\pagerange{\pageref{firstpage}--\pageref{lastpage}}
\maketitle

\begin{abstract}
We present results of a search for extremely metal-poor (EMP) stars in the Large Magellanic Cloud, which can provide crucial information about the properties of the first stars as well as on the formation conditions prevalent during the earliest stages of star formation in dwarf galaxies. Our search utilised SkyMapper photometry, together with parallax and proper motion cuts (from Gaia), colour-magnitude cuts (by selecting the red giant branch region) and finally a metallicity-sensitive cut. Low-resolution spectra of a sample of photometric candidates were taken using the ANU 2.3m telescope/WiFeS spectrograph, from which 7 stars with $\rm [Fe/H] \leq -2.75$ were identified, two of which have $\rm [Fe/H] \leq -3$. Radial velocities, derived from the CaII triplet lines, closely match the outer rotation curve of the LMC for the majority of the candidates in our sample. Therefore, our targets are robustly members of the LMC based on their 6D phase-space information (coordinates, spectrophotometric distance, proper motions and radial velocities), and they constitute the most metal-poor stars so far discovered in this galaxy.
\end{abstract}




\section{Introduction}

The oldest stars in the Universe are important chemical specimens as they can potentially shed light on the formation and evolution of galaxies during the earliest epochs of star formation. One way to search for these very old stars is by looking for extremely metal-poor (EMP) stars ($\rm [Fe/H] \leq -3.0$) amongst the field populations of the Milky Way and of nearby galaxies. These second-generation stars with extremely low abundances of elements (e.g. Fe) were formed out of the gas enriched by the supernovae of their metal-free predecessors, and are able to live long enough to be observable at the present day due to their low mass. Thus, EMP stars can provide us with important, otherwise inaccessible information about the properties of some of the first stars in the Universe, as well as about the initial conditions of star formation found in young galaxies. Moreover, in the Milky Way, we can investigate how some of these stars may have entered the Galactic halo through the accretion of their original host system \citep[see, for example, the review of][]{Frebel2015}.

EMP stars have also been found in dwarf galaxy satellites of the Milky Way. However, such stars are often less well studied than EMP stars in the Galactic halo mainly because the large distances (20-200 kpc) of their host systems means that only the brightest red giants can currently be observed at sufficiently high signal-to-noise with high-resolution spectrographs \citep[e.g.][]{Frebel2015}. An example of this can be found in \citet{Skuladottir2021}, where EMP stars with metallicities down to $\rm [Fe/H]=-4.11$ have been found in the Sculptor dwarf spheroidal galaxy. The EMP stars in dwarf galaxies are generally similar in terms of the element abundance ratios when compared to the Milky Way halo. However, there appears to be some interesting trends in key elements that appear to correlate with the mass of the galaxy. One example is found in the carbon measurements, whereby a large proportion of C-enhanced stars have been found in the Milky Way halo and ultra-faint dwarf galaxies, but only a handful of such stars have been found in classical dwarf galaxies.
\citep[e.g.][]{Tafelmeyer2010,Ishigaki2014,Mashonkina2017,Spite2018}.
Additionally, while the metallicity range spanned by the bulk of stars in the Milky Way and in all classes of dwarf galaxies is similar (reaching below $\rm [Fe/H] = -4$), the ultra-faint dwarf galaxies seem to lack stars with $\rm [Fe/H] \geq -1.5$ \citep{Chiti2018}.

The Large and Small Magellanic Clouds are the largest Milky Way satellites. Despite their importance to many fields of galactic archaeology \citep[e.g.][]{Nidever2017}, not much is known about their EMP star populations. The lowest metallicities of current Magellanic samples do not reach the same extremely low metallicities seen in either the Milky Way halo or in the less massive dwarf satellites. The most metal-poor star known in the LMC, as reported in the literature, has $\rm [Fe/H] = -2.7$ \citep{Reggiani2021}.
Interestingly, this study also found that the most metal-poor Magellanic stars in their sample are r-process enhanced relative to the Milky Way, implying differences in chemical enrichment processes or timescales. This emerging field of chemical abundance analysis has motivated us to initiate a dedicated, sensitive search for EMP stars in the Magellanic Clouds.

Currently, one of the most efficient ways of searching for EMP stars is by using the SkyMapper telescope, which is conducting an imaging program spanning the entire southern sky in 6 filters \citep{Onken2019}. A unique feature of SkyMapper is its $v$ filter, which allows the identification of stars with $\rm [Fe/H] \leq -2.5$ by means of a metallicity-sensitive colour-colour diagram. This has been used to search for EMP stars in the Milky Way with very high efficiency ($\geq$~40\% of photometrically-selected candidates having $\rm [Fe/H] \leq -2.75$; e.g.,  
\citealt{DaCosta2019, Chiti2020,Yong2021}), leading to the discovery of several ultra-metal poor stars ($\rm [Fe/H] \leq -4.0$), including the lowest ever detected abundance of iron in a star (SMSS J160540.18--144323.1; $\rm [Fe/H] = -6.2$; \citealt{Nordlander2019}) and the most iron-poor star known (SMSS J031300.36--670839.3; $\rm [Fe/H] \leq -6.5$; \citealt{Keller2014}).

In this study, we used SkyMapper photometry to conduct a search for EMP stars across the Magellanic system. In particular, we focused on the LMC as it is closer compared to the SMC, which allows us to obtain better quality data. Our photometric EMP selection imposed parallax and proper motion cuts (from Gaia), colour-magnitude diagram (CMD) cuts (by selecting the RGB region) and finally a metallicity-sensitive cut. We obtained low-resolution spectra of a sample of photometric candidates, and used a spectrophotometric analysis to yield metallicity measurements with a precision of $\leq$0.3 dex. Radial velocities, derived from the CaII triplet lines, have also been determined for all candidates in our sample to compare with the systemic velocity of the LMC.

We present the selection method for the EMP candidates
in Section~\ref{sec:sec2} and our observation, data reduction and spectrophotometric analysis methods in Section~\ref{sec:sec3}. In Section~\ref{sec:4}, we present the abundance results for our EMP candidates, and confirm that they are likely members of the Large Magellanic Cloud.

\section{Photometric selection}
\label{sec:sec2}
\subsection{Initial selection}
The LMC EMP candidates were chosen from the SkyMapper Southern Sky Survey Data Release DR3\footnote{The data can be accessed via \url{https://skymapper.anu.edu.au}}, which consists of photometry for over 600 million objects covering a total area of more than 24,000 $\rm deg^{2}$ across all six SkyMapper optical filters: $u$, $v$, $g$, $r$, $i$, $z$ \citep{Wolf2018,Onken2019}. We focused on a region that is 5--20 degrees in radius from the LMC centre (Equation \ref{eq:angle}) to avoid crowding in the inner parts of the dwarf galaxy.

The initial selection from the DR3 database used similar parameters to the ones used in \citet{DaCosta2019} for consistency. The only difference is in the brightness cut, where $g_{\rm psf} \leq 16.5$ so that the LMC candidates can be observed at high dispersion on 8m-class telescopes with sensible integration times.

In addition, we have an advantage in our selection as we know that our desired targets should be located in the LMC. Therefore, we cross-matched our photometric EMP selection sample with Gaia DR2 \citep{Gaia2018}, updated to DR3 \citep{Gaia2021} after data acquisition, to impose further cuts using parallax and proper motion information. The adopted cuts are shown in Equations \ref{eq:parallax} and \ref{eq:pm}, respectively. In particular, Equation \ref{eq:pm} was determined quantitatively, where we measured the average proper motion and spread for stars located 2\degr--6\degr\ from the LMC centre (to avoid kinematics from the core and contamination from the Milky Way halo). We selected the central 90 percentiles of the proper motion distribution. The cuts we adopted were as follow, 
\begin{equation}
     5\si{\degree} \leq \cos^{-1}(\sin(\delta)\sin(\delta_{c}) + \cos(\delta)\cos(\delta_{c})\cos(\alpha-\alpha_{c})) \leq 20\si{\degree}
	\label{eq:angle}
\end{equation}
where $\delta_{c}=-69.78\si{\degree}$ and $\alpha_{c}=81.28\si{\degree}$,
\begin{equation}
     -0.2\arcsec \leq \pi \leq 0.2\arcsec
	\label{eq:parallax}
\end{equation}
\begin{equation}
     \sqrt{1.7(\mu_{\alpha}-1.80)^{2}+0.8(\mu_{\delta}-0.37)^{2}} \leq 1.0.
	\label{eq:pm}
\end{equation}

We note that the centres of the distributions are offset from the Gaia DR2 mean values as provided in \citet{Gaia2018_2}. This is due to asymmetries in the proper motion distribution.

\subsection{Final adopted selection}
\begin{figure}
    \includegraphics[width=\columnwidth]{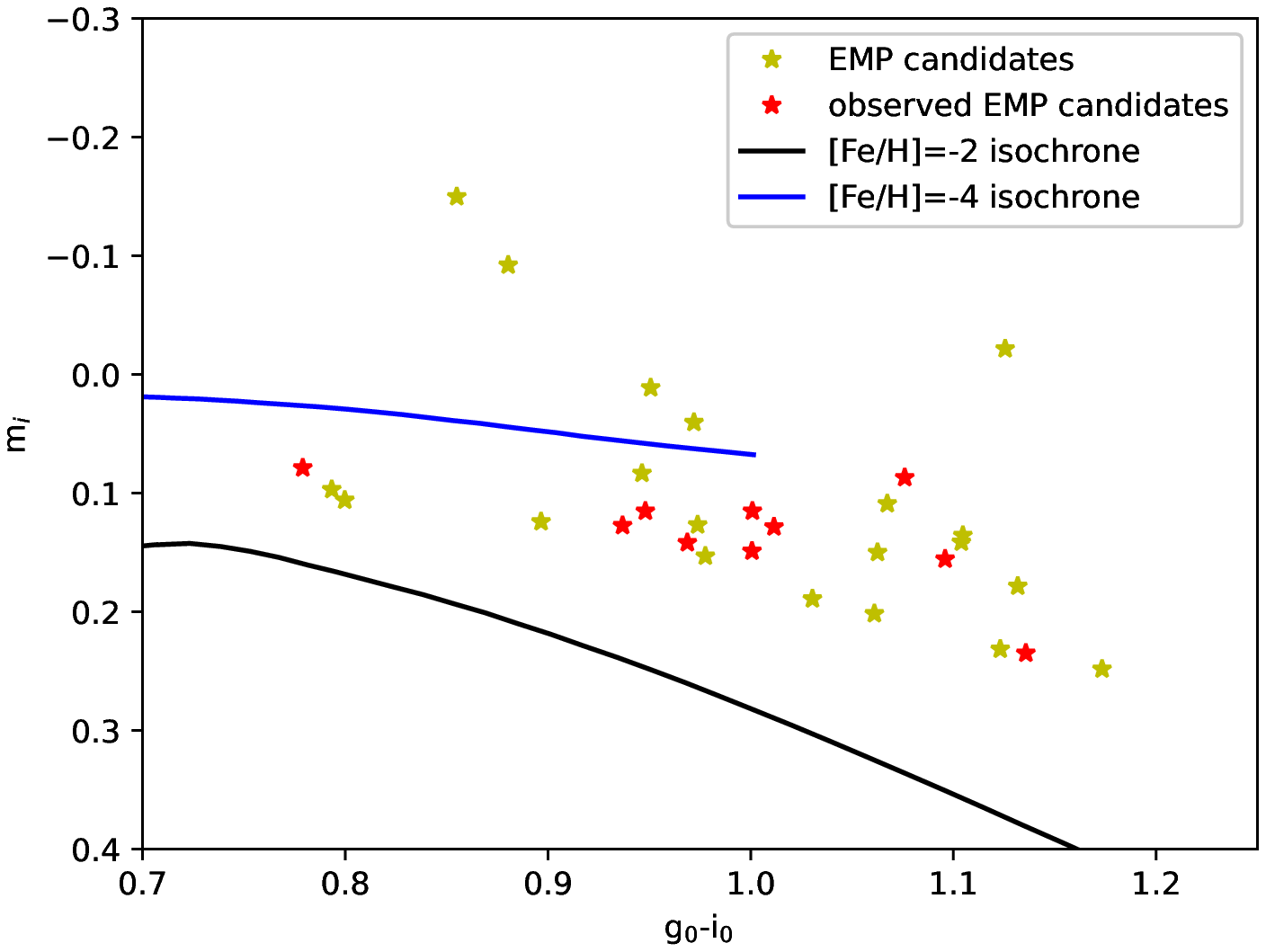}
        \includegraphics[width=\columnwidth]{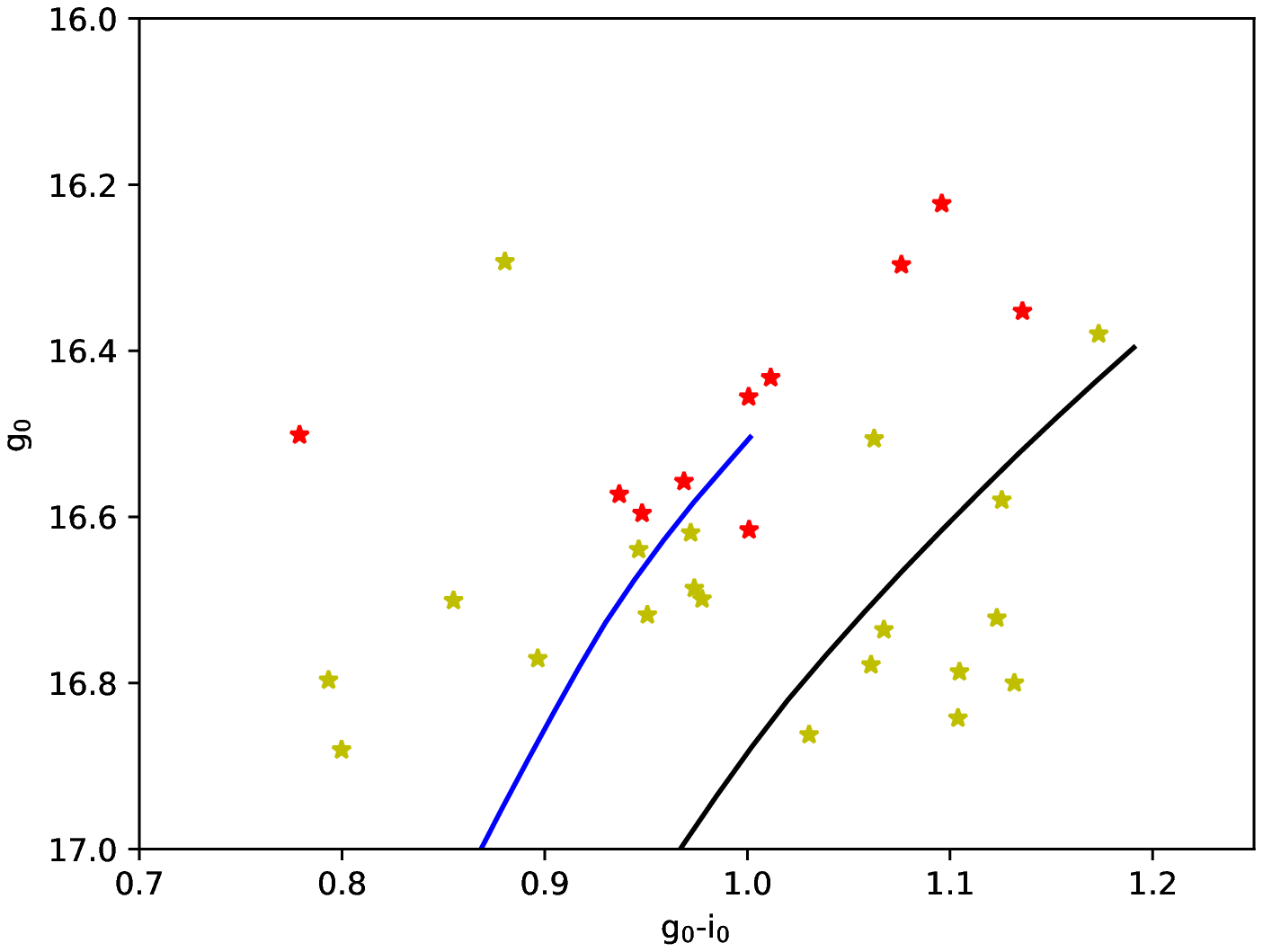}
    \caption{Top: Metallicity-sensitive diagram showing the Magellanic EMP candidates from our final adopted selection. The red stars represent the candidates that have been observed. The blue and black curves represent $\rm [Fe/H] = -4$ and $-2$ Dartmouth isochrones for an age of 12.5 Gyr and $\rm [\alpha/Fe] =0.4$ \citep{Dotter2008}. Both isochrones have been calibrated with reference to the SkyMapper DR3 data as described in \citet{DaCosta2019}. Bottom: CMD with the same description as above. The isochrones have been shifted in distance modulus by 18.5 to match the LMC distance.}
    \label{fig:CMD}
\end{figure}

Given our EMP selection method is a tried and tested method as demonstrated in \citet[see Fig.~3]{DaCosta2019}, we have adopted a similar final selection window in terms of the colour as well as the metallicity index of our candidates (Fig.~\ref{fig:CMD}). In particular, we have selected targets with $0.7 \leq (g-i)_{0} \leq 1.2$ since the RGB at lower metallicities ($\rm [Fe/H] \leq -2.5$) does not go beyond $(g-i)_{0} \approx 1.2$ mag for an old age ($\geq$ $\sim$10 Gyr) population.  The blue colour cut is applied because we want to avoid contamination from young approximately solar metallicity disc dwarfs in the bluer parts of the CMD. As for the metallicity index cut, we have decided to use $m_{i} \geq -0.2$ as previous studies have shown that stars with metallicity indices more negative than this value are often young stars with Ca II H+K emission, or extragalactic objects such as QSOs and active galactic nuclei \citep{DaCosta2019}. The upper (positive) bound on the metallicity index selection was set by the location of the $\rm [Fe/H] = -2$ Dartmouth isochrone for an age of 12.5 Gyr and $\rm[\alpha/Fe] =0.4$ \citep{Dotter2008} in this colour-colour diagram.

\begin{figure}
    \includegraphics[width=\columnwidth]{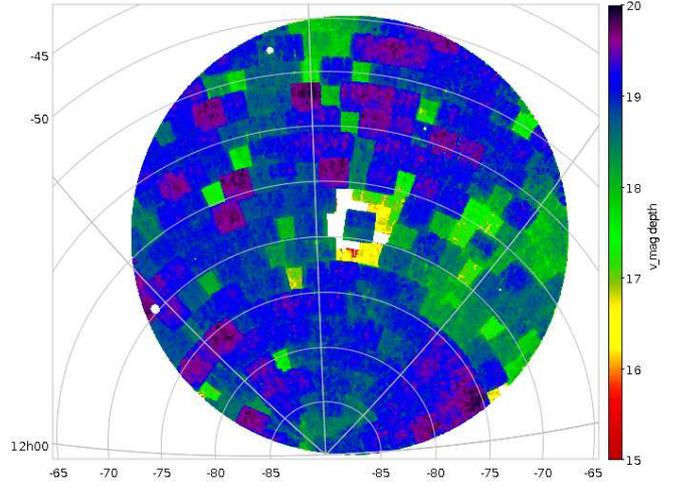}
    \caption{The SkyMapper DR3 $v$-filter coverage in the vicinity of the LMC (20 degree radius). The darker patches show areas with deeper coverage.}
    \label{fig:LMC_vband}
\end{figure}

From a total of around 20,000 candidates in our initial selection, our final adopted selection trims the list down to just 30 candidates. Candidates were prioritised for observing based on their brightness, their relative distance to the $\rm [Fe/H]=-4$ isochrone in the metallicity-sensitive diagram and on their location in the CMD as shown in Fig. \ref{fig:CMD}. 
As shown in Fig.~\ref{fig:LMC_vband}, the depth of the DR3 $v$-band coverage within our annular selection region is not by any means uniform. In particular, the $v$-band coverage in terms of depth reached for e\_v\_psf $<$ 0.2 mag is significantly worse on the western side of the selection region, particularly for $\alpha \leq 4$~hrs and $-75\degr \leq \delta \leq -57\degr$ (Fig. \ref{fig:LMC_vband}). There are also patches of sky with similar reduced $v$-band depth scattered across the entire region. Consequently, we cannot in any sense claim to have generated a complete sample of LMC EMP candidates. 

\section{Observations and Data Reduction}
\label{sec:sec3}

\subsection{Observation using ANU 2.3m/WiFeS}
\label{sec:obs}

Spectroscopic observations of our EMP candidate stars were obtained with the ANU 2.3m telescope at Siding Spring Observatory over twelve nights, spanning from November 2020 to December 2021. Due to bad weather, useful data were obtained on about 60 percent of the allocated nights.  As a result we were able to successfully observe only 10 of the 30 candidates. 

The 2.3m observations were conducted with the WiFeS integral field (IFU) spectrograph \citep{Dopita2010} using the B3000 and I7000 gratings, which cover the wavelength intervals 3200--5900\,\AA\ and 6830--9120\,\AA, respectively. The number behind the letter represents the spectral resolution of the gratings. Multiple 1800\,s exposures were used to yield a signal-to-noise ratio (S/N) per pix in the final summed spectra of $\geq20$ for the blue spectra at the H and K lines of Ca II, and $\geq30$ for the red spectra at the Ca~II triplet lines. Because WiFeS is an integral field spectrograph, useful spectra were still obtained in poor-seeing conditions.

\subsection{Spectrophotometric fits}
\label{sec:fits}

Similar to the process described in \citet{DaCosta2023}, the raw observed spectra of each candidate were sky-subtracted, wavelength-calibrated, combined, and crucially, flux-calibrated using observations of several flux standards each night. These standards are found in \citet{Bohlin2014}, and have well established flux distributions at optical wavelengths. 

The flux-calibrated spectra are then compared to an extension of the grid of model fluxes discussed in \citet{Nordlander2019} using the FITTER code \citep{Norris2013} to determine the best fit. The grid of model spectra is interpolated to a final grid step of 25\,K in $T_{\rm eff}$, and 0.125\,dex in $\log\,g$ and $\rm [Fe/H]$, and the results are rounded off to a maximum of 2 decimal places.

Given that the EMP candidates are assumed to be in the LMC, we utilised the LMC distance to calculate fundamental $\logg$ values which we can then compare to the $\logg$ values obtained from the spectrophotometric analysis. The fundamental $\logg$ calculation is shown by Equation~\ref{eq:gravity} below:
\begin{equation}
    \log\left(\frac{g}{g_\odot}\right)=\log\left(\frac{M}{M_\odot}\right)+4\log\left(\frac{T_\textrm{eff}}{T_\odot}\right)+0.4(M_\textrm{bol}-M_{\textrm{bol}\odot}).
	\label{eq:gravity}
\end{equation}

The steps we took were: assuming masses of $\sim 0.8\Msol$ for the EMP candidates, calculating the bolometric magnitudes using the V magnitudes (which was converted from G and ($G_{bp}-G_{rp}$) from Gaia DR2), bolometric corrections described in \citet{Alonso1999}, using LMC reddening information from \citet{Skowron2021} and assuming the LMC distance modulus to be $18.52 \pm 0.1$ mag \citep{Kovacs2000}. The solar bolometric reference value and effective temperature were taken to be 4.74 \citep{Mamajek2015} and 5770 K respectively. The effective temperature ($\Teff$) estimates used were taken from the spectrophotometric analysis as described above. 
Our results show that the fundamental and spectrophotometric $\logg$ values display a mean offset (fundamental minus spectrophotometric) of 0.11 $\pm$ 0.03 (standard error of mean) with a standard deviation of $\sigma = 0.23$\,dex. The dispersion value is comparable to the spectrophotometric $\logg$ uncertainties discussed in \citep{DaCosta2019}.
The overall agreement between the log $g$ values calculated assuming the stars are at the distance of the LMC, and the distance independent spectrophotometric values, indicates that the observed stars are indeed likely LMC members.

Apart from the EMP candidates, stars with known metallicities from high-resolution spectroscopic studies were also observed and analysed in the same way as our targets to verify the accuracy and precision of our measurements. The outcome is shown in Table \ref{tab:feh_comparison}. The $\Teff$, $\logg$ and $\FeH$ values determined for these stars were compared with the ones obtained from the literature, and the mean offsets, standard error of the mean and sigma values for each stellar parameter were calculated to be $181 \pm 68$K ($\sigma = 84 K$), $0.10 \pm 0.04$ ($\sigma = 0.30$) and $0.20\pm 0.07$ ($\sigma = 0.38$) respectively. We note that the reference temperatures are derived either directly from photometric methods (\citealt{Barklem2005}; \citealt{Hansen2013}), or from spectroscopic methods with corrections to achieve agreement with the photometric temperature scales \citep{Jacobson2015}. Nevertheless, the dispersion values are still comparable to the ones found in \citet{DaCosta2019}.

\subsection{Radial velocities}

The Ca II triplet spectral region (8580-8720\,\AA) was used to measure the radial velocities of the stars. This was done using fxcor in IRAF with two NGC 7099 stars (S1 \& 12917) as templates. We then evaluated the zero point of our velocities by deriving radial velocities for the metallicity reference stars and comparing the velocities with their known values. The mean offset (measured minus reference) and dispersion are $-0.1$ \kms\ and 3.8 \kms\ respectively, indicating that the zero point uncertainty in our velocities is small. The uncertainties in the candidate velocities range from 2--11 \kms, which is derived from the  fxcor velocity error. The radial velocities of our sample were then compared to the LMC rotation curve to confirm their Magellanic membership, which will be elaborated more in Section \ref{sec:4.1}.

\section{Results and Discussion}
\label{sec:4}

\subsection{Magellanic membership} 
\label{sec:4.1}
\begin{figure}
	\includegraphics[width=\columnwidth]{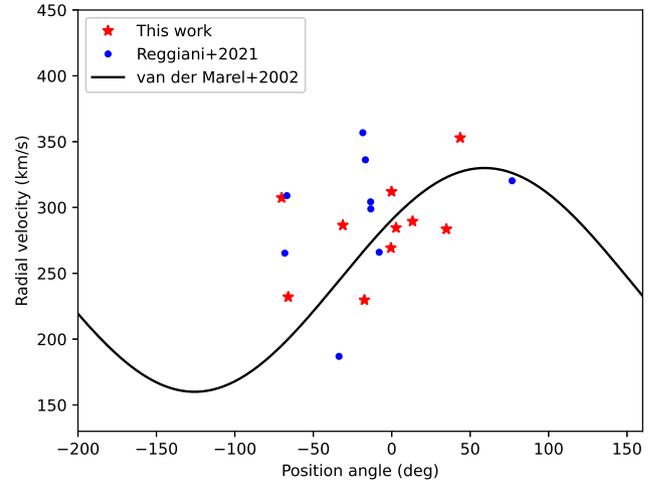}
    \caption{Radial velocity as a function of position angle in the LMC (measured east of north) for our LMC sample. The LMC targets from \citet{Reggiani2021} have also been plotted for reference. The black sinusoid shows the LMC rotation model derived from carbon stars in the outer LMC disk \citep{vandermarel2002} that sit at radii similar to the majority of our EMP sample.}
    \label{fig:rv}
\end{figure}

The radial velocity measurements of our EMP candidates were used to investigate LMC membership. As shown in Fig. \ref{fig:rv}, our candidates' radial velocity measurements follow the rotation curve of the outer parts of the LMC (9\si{\degree}--13\si{\degree} radius; \citealt{vandermarel2002}), which is appropriate given that our targets lie in the LMC outskirts. The agreement is quantified by the results given in Table \ref{tab:rv_std}, which compares the standard deviation of the radial velocities to the standard deviation of the residuals (radial velocity minus rotation curve). The equivalent results for the LMC members discussed in \citet{Reggiani2021} are also given in the Table. These dispersion values are comparable to the dispersion values of 30-40 \kms\ in the LMC outskirts given in \citet{vandermarel2002}, which are based on radial velocities for $\sim$20 LMC carbon-star members.

We note that one star in our sample, 499901368, shows a larger divergence from the rotation curve (123 \kms), and is also the star that lies substantially to the blue of the RGB in the CMD shown in the lower panel of Fig.\ \ref{fig:CMD}. At 19.3\si{\degree}, this star is also the one with the largest angular distance from the centre of the LMC\@. However, it is currently unknown if the disk rotation model still applies for such large distances from the LMC centre. \citet{Cullinane2022} showed that the outer disk of the LMC is highly perturbed, leading to significant spreads in both proper motions and radial velocities (refer to Table 1 of that paper). Therefore, given that the star's radial velocity measurement is only $\sim40$ \kms\ higher than the systemic velocity of the LMC ($\sim262$ \kms; \citealt{vandermarel2002}), we are still confident that 499901368 is a member of the LMC.

Therefore, we can conclude that our candidates are likely to be LMC members due to following reasons as summarised: Targets match the systemic LMC circular velocity at their position angle; Proper motions are consistent with that of the LMC; $\logg$ values calculated under the assumption of lying at the LMC distance agree with distance-independent spectrophotometric $\logg$ values.

\begin{table}
	\centering
	\caption{Comparing the mean radial velocity rv, its standard deviation, and the residual dispersion $\sigma_{\rm res}$ (radial velocity minus rotation curve), for our sample and that of \citet{Reggiani2021}. The number of stars in each sample is also given.}
	\label{tab:rv_std}
	\begin{tabular}{lccc} 
		\hline
		Sample & N & rv (\kms) & $\sigma_{\rm res}$ (\kms)\\
            \hline
            This work & 10 & $284.8\pm34.6$ & 43.5\\
            Reggiani et al.\ (2021) & 9 & $293.7\pm47.0$ & 52.4\\
		\hline
	\end{tabular}
\end{table}

\subsection{Stellar metallicities}

\begin{figure}
	\includegraphics[width=\columnwidth]{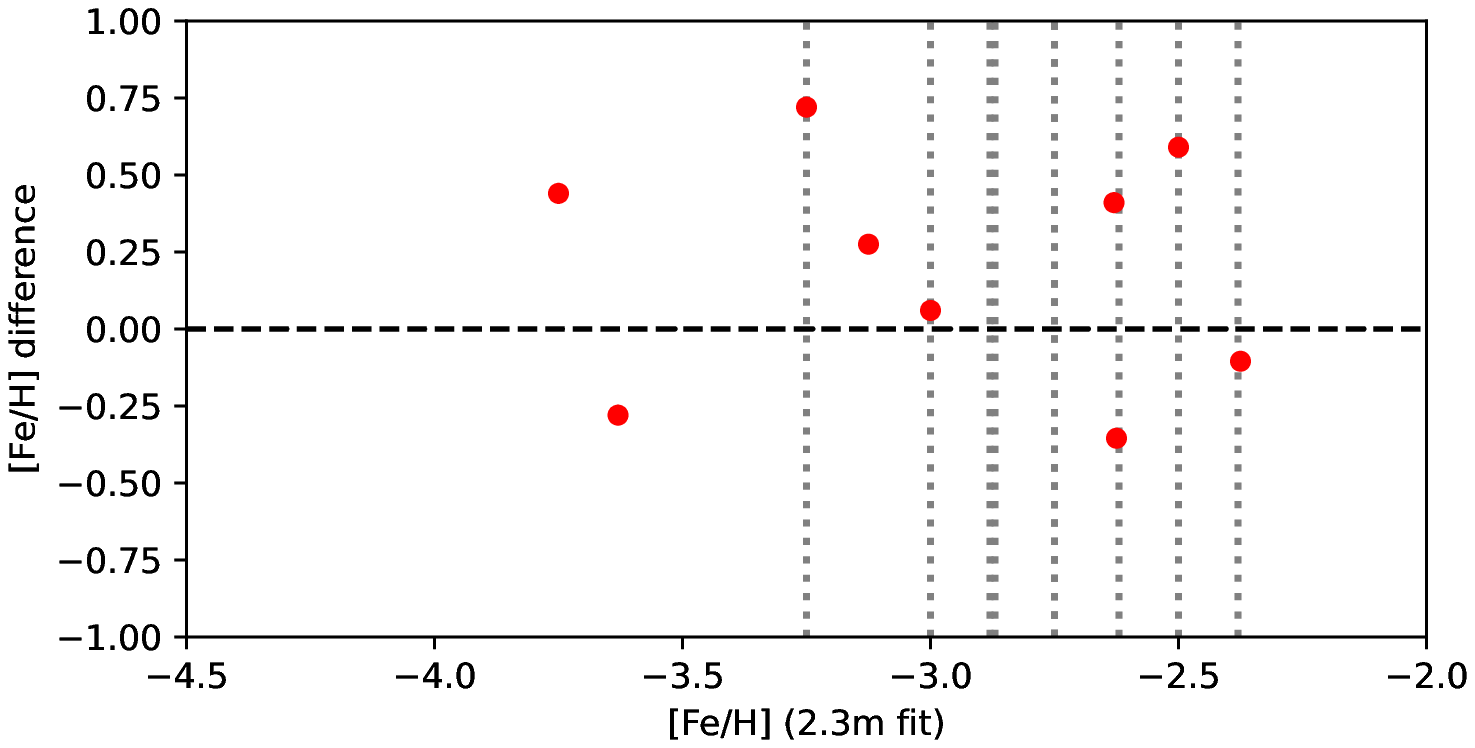}
        \includegraphics[width=\columnwidth]{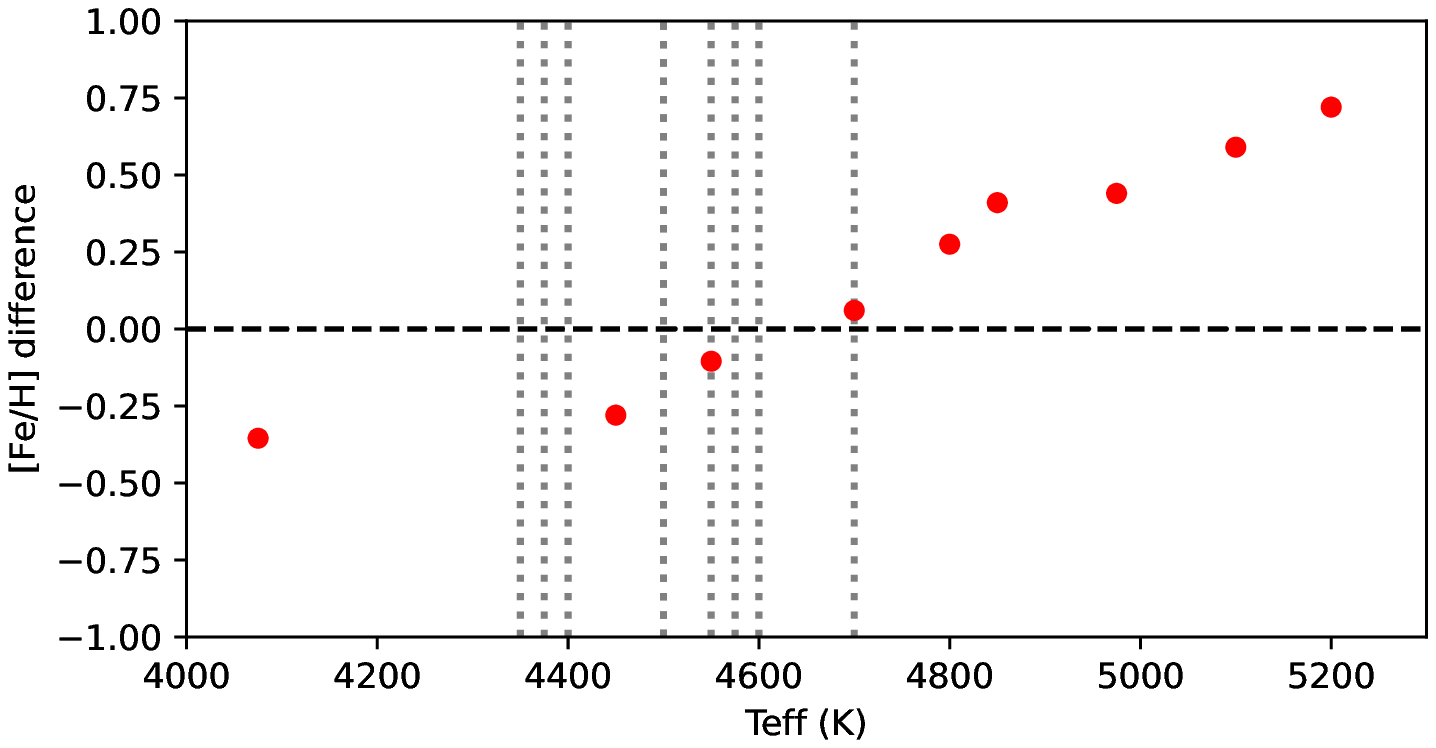}
    \caption{Differences in [Fe/H] between our measurements and the literature vs [Fe/H] (top panel) and $\Teff$ (bottom panel) for the reference stars. The vertical dotted lines indicate the stellar parameters of our sample.}
    \label{fig:feh_diff}
\end{figure}

As mentioned in Section \ref{sec:fits}, we observed a set of stars with known metallicities from high dispersion analyses to verify the accuracy of our metallicity analysis. Overall, our low-resolution spectrophotometric measurements are comparable with that of the literature as shown in Table \ref{tab:feh_comparison} and Figure \ref{fig:feh_diff}. We found a mean $\FeH$ difference (measured minus reference) of $0.20 \pm 0.07$\,dex ($\sigma = 0.38$\,dex). This shows that the metallicity results for our EMP candidates are consistent. The level of uncertainty is similar to that listed in \citet{DaCosta2019} and \citet{Yong2021}, where the spectrophotometric metallicities are shown to have uncertainties at the $\pm$0.3 dex level. 

\begin{figure*}
	\includegraphics[width=\columnwidth]       {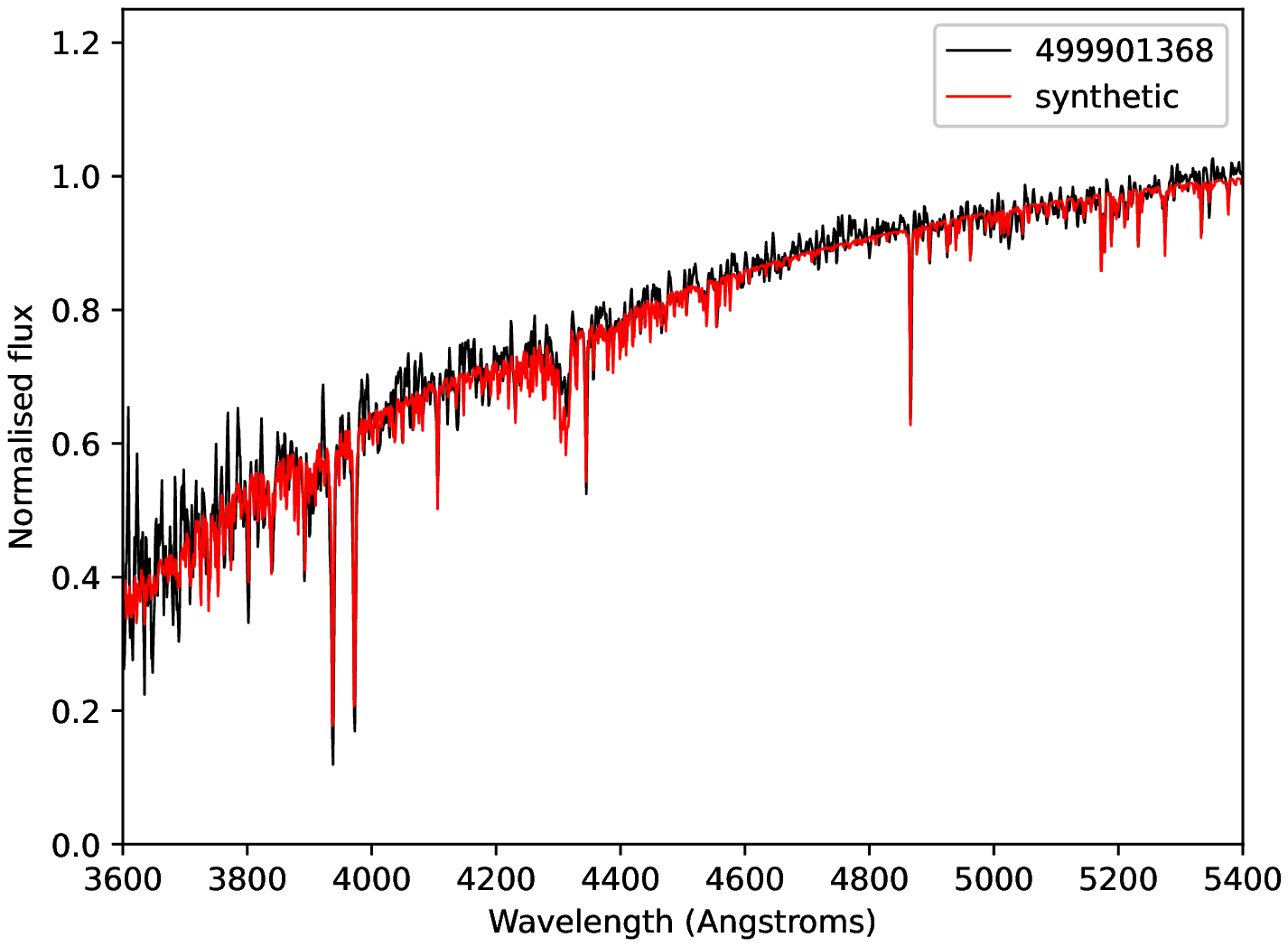}
        \includegraphics[width=\columnwidth]
        {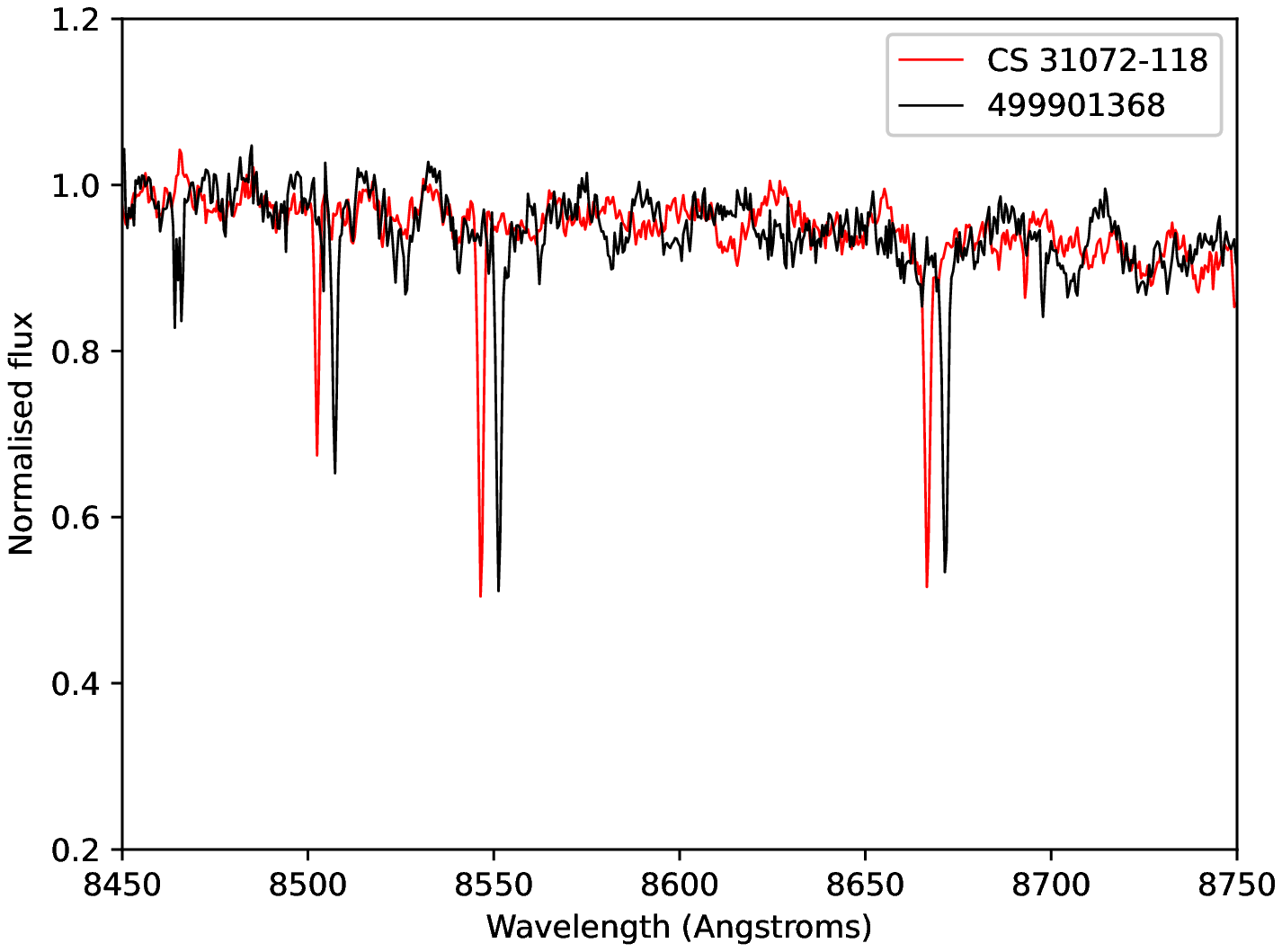}
        \caption{\textit{Left}: The B3000 spectrum (black line) from the 2.3m WiFeS observations of the most metal-poor EMP candidate 499901368.  The relative flux values have been normalized to unity at 5500\,\AA. The spectrophotometric fitting process yields 4700/1.63/$-$3.25 for $\Teff$, $\logg$ and [Fe/H], respectively, and the best-fitting model spectrum is overplotted in red. \textit{Right}: The I7000 spectrum (black line) of the same star as above, showing the Ca II triplet region. The spectrum of reference star CS 31072-118  ($\rm[Fe/H]_{lit}=-3.06$) is overplotted in red. The spectra have not been corrected for radial velocities. Our spectrophotometric analysis of the B3000 spectrum indicates similar stellar parameters for both stars (CS 31072-118: 4700/1.00/$-$3.00). The close similarity of the Ca II triplet line strengths in the two observed spectra confirms the similarity of the metallicities. }
    \label{fig:spectra}
\end{figure*}

Table~\ref{tab:stellar_param} shows the stellar parameters of the Magellanic EMP candidates, including their metallicities, from the spectrophotometric fits. Also given are the coordinates, radius from the LMC centre, position angle, reddening, photometry and radial velocity information.
We have identified 7 stars with $\rm [Fe/H] \leq -2.75$, of which two have $\rm [Fe/H] \leq -3$. The B3000 and I7000 spectra of the most metal-poor star in our sample (499901368: $\rm [Fe/H] = -3.25$) are shown in Fig. \ref{fig:spectra}, with the left panel also showing the spectrophotometric fit for this star. The right panel compares the spectrum of the candidate with that for CS~31072-118 which has similar stellar parameters. Our sample thus reveals the most metal-poor stars so far discovered in the LMC.

\subsection{Selection efficiency}

Of the 10 EMP candidates that we observed, 7 (or 70\%) have $\rm [Fe/H] \leq -2.75$, while 2 (or 20\%) have $\rm [Fe/H] \leq -3.00$. Overall, all of the stars observed have $\rm [Fe/H] \leq -2.00$. These numbers are comparable to the ones found in \citet{DaCosta2019}, and they confirm the high selection efficiency of SkyMapper photometry for identifying candidate metal-poor stars.

\subsection{Carbon abundances}

By inspecting the G-band (CH) region of the spectrophotometric fits to the blue spectra, which employ [C/Fe] = 0, we find no evidence of carbon enhancement in any of our EMP candidates. This is similar to the results found in \citet{DaCosta2019}, which determined that the selection of metal-poor stars based on SkyMapper filters may be biased against highly carbon-rich EMP stars.  As discussed in \citet{DaCosta2019}, this is most likely due to large carbon enhancements affecting the metallicity index, making CEMP objects appear to be more metal-rich, and hence not selected as potential EMP candidates.

\begin{table*}
	\centering
	\caption{Comparing stellar parameters of the reference stars from our study with that of the literature. The reference stellar parameters are provided by 1. \citet{Barklem2005}, 2. \citet{Hansen2013}, 3. \citet{Jacobson2015} and 4. \citet{Harris1996}.}
	\label{tab:feh_comparison}
	\begin{tabular}{lccccrrrrrr} 
		\hline
		Object & $\rm{T}_{\textrm{eff,low}}$ & $\rm{T}_{\textrm{eff,lit}}$ & $\Delta \rm\Teff$ & $\logg_\textrm{low}$ & $\logg_\textrm{lit}$ &  $\Delta \logg$ & $\rm[Fe/H]_{low}$ & $\rm[Fe/H]_{lit}$ & $\Delta \FeH$ & Ref\\
		\hline
		HE 0008-3842 & 4450 & 4327 & 123 & 1.00 & 0.65 & $-$0.35 & $-$3.63 & $-$3.35 & $-$0.28 & 1\\
		CS 31072-118 & 4700	& 4606 & 94	& 1.00	& 1.25 & 0.25 & $-$3.00 & $-$3.06 & 0.06 & 1\\
            CS 29491-053 & 4850	& 4700 & 150 & 1.00 & 1.30	& 0.30 & $-$2.63 & $-$3.04 & 0.41 & 2\\
            SMSS J022410.38$-$534659.9 & 4800 & 4630 & 170 & 1.13 & 0.90 & $-$0.23 & $-$3.13 & $-$3.40 & 0.27 & 3\\
            SMSS J040148.04$-$743537.3 & 5100 & 4900 & 200 & 1.88 & 1.90 & 0.02 & $-$2.50 & $-$3.09 & 0.59 & 3\\
            SMSS J022423.27$-$573705.1 & 5200 & 4846 & 354 & 1.13 & 1.60 & 0.47 & $-$3.25 & $-$3.97 & 0.72 & 3\\
            CD $-38^{\circ}\,245$ & 4975 & 4800 & 175 & 1.25 & 1.50 & 0.25 & $-$3.75 & $-$4.19 & 0.44 & 2\\
            NGC 7099-12917	& 4075 & - & - & 1.00 & - & - & $-$2.63 &$-$2.27 & $-$0.36 & 4\\
            NGC 7099-S1	& 4550 & - & - & 1.00 & - & - & $-$2.38 & $-$2.27 & $-$0.11 & 4\\
		\hline
	\end{tabular}
\end{table*}

\begin{table*}
	\centering
	\caption{Complete table showing full spectrophotometric information of the LMC EMP candidates. The radius is given relative to the LMC centre at $(\alpha,\delta)=(81.28\si{\degree},-69.78\si{\degree})$, and the position angle is measured east of north. The photometric data was retrieved from SkyMapper DR3, including the metallicity-sensitive index $m_i = (v-g)_0 -1.5(g-i)_0$, and was dereddened using E(B-V) derived from OGLE-IV reddening maps \citep{Skowron2021}. Stellar parameters were derived from our spectrophotometric analysis.}
	\label{tab:stellar_param}
	\begin{tabular}{lllrrccccccccc} 
		\hline
		SMSS DR3 & RA & Dec. & Radius & Angle & $g_{0}$ & $(g-i)_{0}$ & E(B-V) & $m_{i}$ & $\Teff$ (K) & $\logg$ & [Fe/H] & rv (\kms)\\
		\hline
            500263380 & 05 24 59.3 & $-$62 48 07.4 & 7.0\si{\degree} &	$-$0.1\si{\degree} & 16.352 & 1.136 & 0.052 & 0.235 & 4350	& 0.50	& $-$2.38	& $312.0\pm2.2$\\
            471817979 & 06 29 57.1 & $-$60 24 33.3 & 11.5\si{\degree} &	43.7\si{\degree} & 16.296 & 1.076 & 0.070 & 0.087 & 4375	& 0.50	& $-$2.50	& $352.9\pm2.5$\\
            500206265 & 05 43 35.5 & $-$59 50 43.2 & 10.1\si{\degree} &	13.3\si{\degree} & 16.432 & 1.001 & 0.059 & 0.149 & 4400 & 1.00	& $-$2.62	& $289.5\pm7.7$\\
            471915910 & 06 08 14.9 & $-$62 07 23.5 & 8.8\si{\degree} &	34.9\si{\degree} & 16.600 & 0.948 & 0.049 & 0.115 & 4575 & 1.00	& $-$2.75	& $283.6\pm6.9$\\
            497519424 & 04 54 53.6 & $-$64 07 52.2 & 6.4\si{\degree} &	$-$31.2\si{\degree} & 16.616 & 1.001 & 0.036 & 0.115 & 4500	& 1.00	& $-$2.75	& $286.5\pm9.6$\\
		500382880 & 05 24 24.1 & $-$59 16 05.5 & 10.5\si{\degree} &	$-$0.5\si{\degree} & 16.456 & 1.001 & 0.032 & 0.149 & 4550 & 1.00 & $-$2.75 &	$269.3\pm6.3$\\
            497682788 & 04 03 47.0 & $-$64 30 55.7 & 9.4\si{\degree} &	$-$66.0\si{\degree} & 16.557 & 0.969 & 0.047 & 0.142 & 4500 & 1.00 &	$-$2.87 &	$232.1\pm8.1$\\
            500287810 & 05 28 29.6 & $-$61 26 49.6 & 8.3\si{\degree} &	2.8\si{\degree} & 16.223 & 1.096 & 0.050 & 0.155 & 4500	& 0.50	& $-$2.88	& $284.6\pm2.6$\\
            500766372 & 04 59 15.4 & $-$58 54 16.0 & 11.2\si{\degree} &	$-$17.4\si{\degree} & 16.573 & 0.937 & 0.023 & 0.127 &	4600 & 1.00	& $-$3.00	& $229.7\pm6.8$\\
            499901368 & 03 02 07.9 & $-$57 53 20.2 & 19.3\si{\degree} &	$-$70.2\si{\degree} & 16.501 & 0.779 & 0.015 & 0.079 & 4700	& 1.63	& $-$3.25	& $307.4\pm3.0$\\

		\hline 
	\end{tabular}
\end{table*}

\section{Conclusions}

We present results of our search for extremely metal-poor (EMP) stars in the Large Magellanic Cloud via SkyMapper photometry. Our photometric selection of EMP stars involved applying cuts on parallax and proper motion (from Gaia), on the color-magnitude diagram (by selecting the red giant branch region), and a metallicity-sensitive cut. To confirm the EMP status of our photometric candidates, we obtained low-resolution spectra using the ANU 2.3m telescope/WiFeS spectrograph combination. We identified seven stars with $\rm[Fe/H] \leq -2.75$, including two with $\rm[Fe/H] \leq -3$. Radial velocities, derived from the CaII triplet lines, generally match well with the outer rotation curve of the LMC for the candidates in our sample, confirming that our targets are probable LMC members. Our results constitute the most metal-poor stars found so far in the LMC. We have obtained high-resolution spectra of the most metal-poor sample and will present our findings in a future paper.

\section*{Acknowledgements}

We thank Dr.\ Chris Onken (RSAA, ANU) for providing information on the (non)-uniformity of the v-band coverage within our selected sky area. This research was supported by the Australian Research Council Centre of Excellence for All Sky Astrophysics in 3 Dimensions (ASTRO 3D), through project number CE170100013. \\
We also acknowledge the traditional owners of the land on which the Siding Spring Observatory is located, the Gamilaraay people, and pay our respects to elders past, present, and emerging.

\section*{Data Availability}

The underlying SkyMapper and ANU 2.3m/WiFeS data will be made available on reasonable request to the authors.



\bibliographystyle{mnras}
\bibliography{EMP} 





\bsp	
\label{lastpage}
\end{document}